\documentclass[iop,apj]{emulateapj}
\usepackage{graphicx}
\usepackage{amsmath}

\def\stacksymbols #1#2#3#4{\def\theguybelow{#2}
        \def\verticalposition{\lower#3pt}
        \def\spacingwithinsymbol{\baselineskip0pt\lineskip#4pt}
        \mathrel{\mathpalette\intermediary#1}}
\def\intermediary #1#2{\verticalposition\vbox{\spacingwithinsymbol
        \everycr={}\tabskip0pt
        \halign{$\mathsurround0pt#1\hfil##\hfil$\crcr#2\crcr
                \theguybelow\crcr}}}

\shorttitle{Metal-rich Trailing Outflows}
\shortauthors{DUAN \& GUO}

\begin{document}
\bibliographystyle{apj} 

\title{Metal-rich Trailing Outflows Uplifted by AGN Bubbles in Galaxy Clusters}

\author{Xiaodong Duan$^{1,2}$ and Fulai Guo$^{1,2*}$}

\affil{$^{1}$Key Laboratory for Research in Galaxies and Cosmology, Shanghai Astronomical Observatory, Chinese Academy of Sciences,
80 Nandan Road, Shanghai 200030, China}

\affil{$^{2}$School of Astronomy and Space Science, University of Chinese Academy of Sciences, 19A Yuquan Road, 100049, Beijing, China}

\altaffiltext{*}{Email: fulai@shao.ac.cn}

\begin{abstract}

Recent {\it Chandra} X-ray observations of many galaxy clusters find evidence for hot metal-rich outflows preferentially aligned with the large-scale axes of X-ray cavities with typical outflow masses of around $10^{9} $ - $10^{10} M_{\odot}$. Here we perform a suite of three hydrodynamic simulations to investigate whether AGN jets could drive these metal-rich outflows in a representative cluster. By using both the tracer variable and virtual particle methods, and additionally following the gas metallicity evolution, we show that metal-rich gas initially located in central regions can indeed be uplifted by the AGN bubble to large distances, a phenomenon called {\it Darwin drift} in fluid mechanics, and forming a filamentary trailing outflow extending beyond $100$ kpc behind the bubble. The gas entrained in the trailing outflow is entirely outflowing with an average outflow rate of nearly $100M_{\odot}$/yr during the first $100$ Myr, and at later times, a growing lower part flows back towards the cluster center due to gravity. The outflow mass rises up to about $10^{10} M_{\odot}$ with entrained iron mass of about $10^{6} - 10^{7}M_{\odot}$, consistent with observations and predictions from the drift model. By the end of our simulation ($\sim 800$ Myr after the AGN event), several $10^{9}M_{\odot}$ of the uplifted high-metallicity gas still remains at large altitudes, potentially contributing to the enrichment of the bulk ICM and the broadening of central metallicity peaks observed in cool core clusters.    
 
\end{abstract}

%%%%%%%%%%%%%%%%%%%%%%%%%%%%%%%%%%%%%%%%%%%%%%%%%%%%%%%%%%%%%%%%%%%%%%%%
%Section1
\section{Introduction}
\label{section1}

Recent {\it Chandra X-ray Observatory} observations of many galaxy clusters reveal anisotropic gas metallicity distributions preferentially elongated along the large-scale axes of radio bubbles and X-ray cavities, suggesting the existence of hot metal-rich outflows driven by AGN jets in the intracluster medium (ICM; \citealt{kirp09}; \citealt{simionescu09}; \citealt{kirp11}; \citealt{kirp15}). The outflow mass ranges typically from $10^{9}$ - $10^{10}M_{\odot}$ with the mean outflow rate of typically tens of solar masses per year and upward of $100M_{\odot}$/yr in the extreme (\citealt{kirp15}). These hot outflows may be physically related to cold molecular outflows with similar outflow masses recently discovered by Atacama Large Millimeter Array observations (e.g., \citealt{mcnamara14}; \citealt{vantyghem16}; \citealt{russell17}), potentially playing an important role in the AGN feedback loop \citep{mcnamara16}.

One way that AGN jets drive anisotropic outflows in galaxy clusters is via the motions of jet-created AGN bubbles in the gravitationally-stratified ICM, which  uplift the ICM gas in their wakes, as previously shown in the simulations of buoyantly-rising AGN bubbles (\citealt{churazov01}; \citealt{saxton01}; \citealt{bruggen03}; \citealt{roediger07}; \citealt{revaz08}). To distinguish these outflows from those swept up by AGN-driven shocks, we refer to these outflows as trailing outflows of AGN bubbles in our recent work (\citealt{guo18}, hereafter G18). Trailing outflows belong to a general phenomenon in fluid mechanics known as Darwin drift (\citealt{darwin53}), which has been investigated extensively with both analytical (\citealt{yih85}; \citealt{pushkin13}) and experimental (\citealt{dabiri06}; \citealt{peters16}) methods. The drift model of trailing outflows in galaxy clusters has been previously studied by \citet{pope10}, which suggest that extended trailing outflows behind buoyant AGN bubbles can indeed be formed in real clusters irrespective of turbulence, as the buoyancy length is usually much larger than the Ozmidov scale (the maximum scale of turbulence eddies). \citet{fabian03} argued that trailing outflows may be laminar and turbulence is significantly suppressed in the ICM by viscosity.

Observations find that elongated hot outflows (\citealt{kirp09}; \citealt{kirp15}) and cool filaments (\citealt{fabian03}; \citealt{hatch06}, \citealt{salom06}; \citealt{salom08}) indeed exist between AGN bubbles and the cluster center in some clusters, consistent with the trailing outflow scenario. Trailing outflows may reach and even enter AGN bubbles, as suggested by some recent observations (\citealt{sullivan13}; \citealt{anderson18}). X-ray observations also indicate that heavy elements are often transported beyond the spatial extent of inner X-ray cavities (\citealt{kirp11}; \citealt{kirp15}), suggesting that this is a long-lasting phenomenon and successive AGN outbursts may be triggered by and interact with trailing outflows created by previous AGN events.

In this paper, we perform a series of three representative hydrodynamic simulations to study hot metal-rich trailing outflows in galaxy clusters. In previous simulations (\citealt{churazov01}; \citealt{saxton01}; \citealt{bruggen03}; \citealt{reynolds05};  \citealt{roediger07}; \citealt{revaz08}), trailing outflows are usually uplifted during the ascent of initially-static cavities created manually at the beginning of these simulations, where the creation of shocks and AGN bubbles by AGN jets is neglected. In G18, we directly model the jet-ICM interaction, and investigate complex gasdynamical processes triggered by this interaction, including forward shocks, trailing outflows, rarefaction waves and meridional circulations. Here in this paper, we continue our previous study in G18, and particularly focus on trailing outflows uplifted by AGN bubbles and propose the {\it Darwin drift} model to explain this phenomenon. We adopt the tracer variable and virtual particle methods to track gas motions in trailing outflows, and additionally follow the evolution of an observationally-motivated gas metallicity distribution, demonstrating for the first time that metal-rich outflows indicated by X-ray observations can indeed be physically uplifted by buoyant AGN bubbles in galaxy clusters.

The rest of the paper is organized as follows. We describe basic equations and our methodology in Section \ref{section2}. We study hot trailing outflows with the tracer variable and virtual particle methods in Section \ref{section3.1}, and further investigate in detail in Section \ref{section3.2} the properties and evolution of metal-rich trailing flows in our main run R3, where gas metallicities are followed passively. We present the mass evolution of trailing outflows in run R3 in Section \ref{section3.3}, and compare the outflow mass with that estimated in the {\it Darwin drift} model in Section \ref{section3.4}. Finally, we summarize our main results in Section \ref{section4}.

%%%%%%%%%%%%%%%%%%%%%%%%%%%%%%%%%%%%%%%%%%%%%%%%%%%%%%%%%%%%%%%%%%%%%%%%%
%Section2  
\section{Methodology}
\label{section2}

In this paper, we performed a series of three hydrodynamic simulations, and adopted the simulation presented in G18 as our base simulation (run R1 listed in Table 1). While G18 studies how AGN jets transfer energy to the ICM and provides an overall picture of complex gasdynamical processes associated with AGN feedback events, here we specifically focus on trailing outflows uplifted by AGN bubbles. To this end, we additionally follow the evolution of a tracer fluid and some virtual particles. Motivated by recent observations of relatively high gas metallicities found in trailing outflows (e.g., \citealt{kirp11}; \citealt{kirp15}), we follow the ICM metallicity evolution in run R3. To better compare with observations, we adopt shear viscosity in runs R2 and R3 to help suppress Kelvin-Helmholtz (KH) instabilities. Below we briefly describe our models and simulation setup with a focus on modifications, and refer the reader to G18 for more details. 

%Section2.1
\subsection{Equations and Numerical Setup}
\label{section2.1}

Incorporating shear viscosity and the ICM metallicity evolution, the ICM evolution may be governed by the following hydrodynamic equations:
\begin{eqnarray}
\frac{d \rho}{d t} + \rho \nabla \cdot {\bf v} = 0,\label{hydro1}
\end{eqnarray}
\begin{eqnarray}
\rho \frac{d {\bf v}}{d t} = -\nabla P-\rho \nabla \Phi +\nabla \cdot {\bf \Pi},\label{hydro2}
\end{eqnarray}
\begin{eqnarray}
\frac{\partial e}{\partial t} +\nabla \cdot(e{\bf v})=-P\nabla \cdot {\bf v}+{\bf \Pi}:\nabla {\bf v} - \mathcal{C}
   \rm{ ,}\label{hydro3}
   \end{eqnarray}
\begin{eqnarray}
\frac{d (\rho Z)}{d t} + \rho Z \nabla \cdot {\bf v} = 0,\label{hydro4}
\end{eqnarray}
  \\ \nonumber
\noindent
where $d/dt \equiv \partial/\partial t+{\bf v} \cdot \nabla $ is the Lagrangian time derivative, and {\bf $\Pi$} is the viscous stress tensor (see \citealt{reynolds05}),
\begin{eqnarray}
\Pi_{\rm ij}=\mu_{\rm visc}\left(\frac{\partial v_{\rm i}}{\partial x_{\rm j}}+\frac{\partial v_{\rm j}}{\partial x_{\rm i}}-\frac{2}{3}\delta_{\rm ij}\nabla \cdot {\bf v}\right)
 {\rm .}
   \end{eqnarray}
In the equations above, $\mu_{\rm visc}$ is the dynamic viscosity coefficient, $Z$ is the ICM metallicity in units of the solar metallicity $Z_{\sun}$, $\Phi$ is the gravitational potential, and $\rho$, $v$, $e$, $P$ are the density, velocity, thermal energy density, pressure of the ICM, respectively. The equation of state of the ICM is assumed to be $P=(\gamma-1)e$ with $\gamma=5/3$, while the molecular weight per particle is assumed to be $\mu=0.61$. In the energy Equation (\ref{hydro3}), $\mathcal{C}$ is the energy loss rate per unit volume due to radiative cooling and may be written as $\mathcal{C}=n_{\rm i}n_{\rm e}\Lambda(T,Z)$, where $n_{\rm e}$ is the electron number density, $n_{\rm i}$ is the ion number density, and the cooling function $\Lambda(T, Z)$ is adopted from \citet{sd93}.

Assuming axisymmetry around the jet axis, we solve Equations (\ref{hydro1}) - (\ref{hydro4}) in $(R, z)$ cylindrical coordinates using our own two-dimensional Eulerian code similar to ZEUS 2D \citep{stone92}. The code has been successfully used in many previous studies, e.g., \citet{guo10a}, \citet{guo11}, \citet{guo12}, and G18. The subroutine for the metallicity evolution has been previously used in \citet{guo10b}, where we refer the reader for details. Our implementation of shear viscosity is described in detail in \citet{guo12b}, and is also adopted in \citet{guo15} and \citet{guo16}. Along either $R$ or $z$ directions, the computational grid consists of $800$ equally spaced zones with spatial resolution of $250$ pc out to $200$ kpc plus additional $400$ logarithmically-spaced zones out to $2$ Mpc. As for boundary conditions, we use reflective boundary conditions at inner boundaries ($z=0$, or $R=0$) and outflow boundary conditions at outer boundaries ($z=2$ Mpc, or $R=2$ Mpc). 

For concreteness, we use the cluster Abell 1795 as our default cluster model, but our results are expected to hold generally for galaxy groups and clusters. As described in detail in G18 (also see \citealt{guo14}), the gravitational potential of our cluster is contributed by three components: the dark matter halo ($\Phi_{\text{DM}}$), the central galaxy ($\Phi_ {*}$), and the central SMBH ($\Phi_{\text{BH}}$), and is assumed to be fixed during our simulations. For the initial ICM temperature profile, we adopt an analytic expression as adopted in G18, which provides a reasonably good fit to {\it Chandra} data of A1795 from the inner few kpc to about 1 Mpc (\citealt{vikhlinin06}; \citealt{guo10b}; \citealt{guo14}). The initial ICM density profile is derived from the initial temperature profile and the gravitational potential, assuming hydrostatic equilibrium.

%Table 1
\begin{table}
 \centering
 \begin{minipage}{80mm}
  \renewcommand{\thefootnote}{\thempfootnote} 
  \caption{List of Simulations.}
    \vspace{0.1in}
  \begin{tabular}{@{}lccccc}
  \hline   \hline 
  & $\mu_{\rm visc}$&$Z$&$t_{\rm cc}$\footnote{$t_{\rm cc}$ is the time when the central cooling catastrophe happens. The jet is manually turned on at $t=t_{\rm cc}$, and lasts for $5$ Myr.}\\
         Run&(g cm$^{-1}$ s$^{-1}$)&$(Z_{\sun})$ & (Myr)\\ \hline 
         R1 .................. &$0$&$0.4$& 238\\                 
         R2 .................. & $150$& $0.4$ &238\\
         R3 .................. &$150$& $Z_{\rm A1795} (r)$ \footnote{Here in run R3, we adopt for the initial metallicity profile a radius-dependent analytic fit (Eq. \ref{eqnmet}) to the observed iron abundance distribution in A1795 as in \citet{guo10b}.}&180 \\ 
          \hline
\label{table1}
%   \vspace{-0.2in}
\end{tabular}
\end{minipage}
\end{table}

%Figure 1
\begin{figure*}
\centering
\includegraphics[height=0.34\textheight]{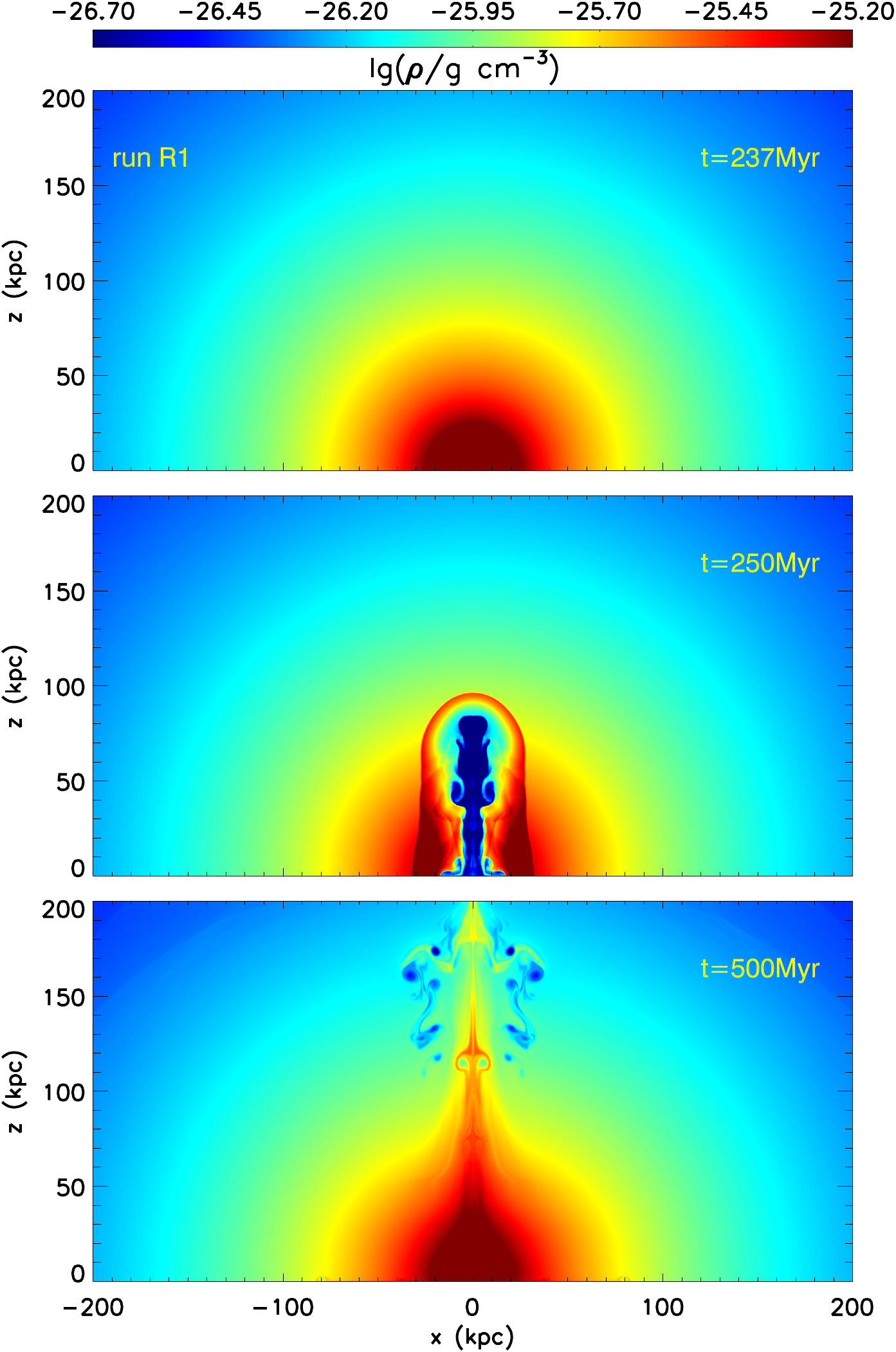}~~~
\includegraphics[height=0.34\textheight]{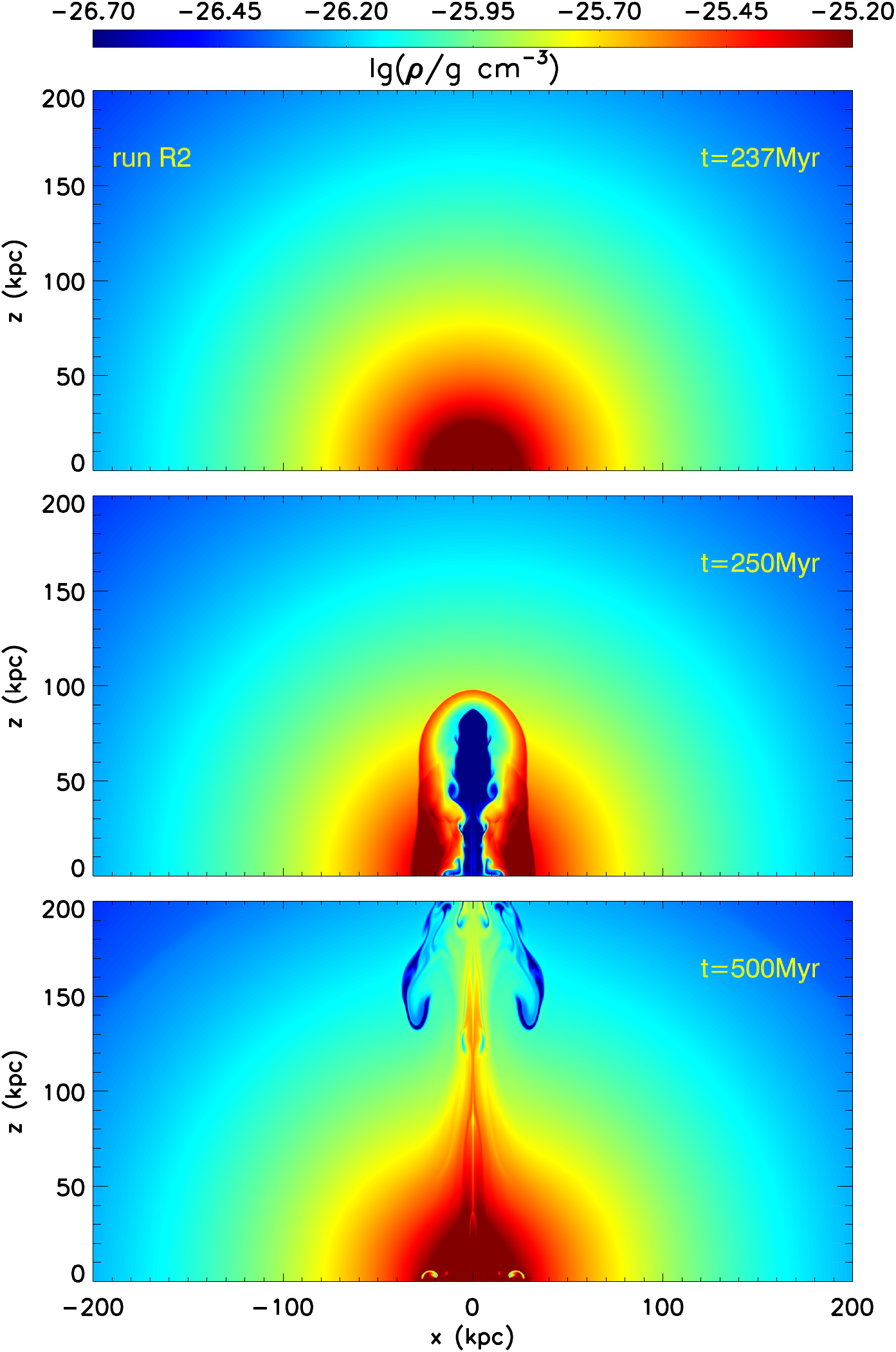}~~~
\includegraphics[height=0.34\textheight]{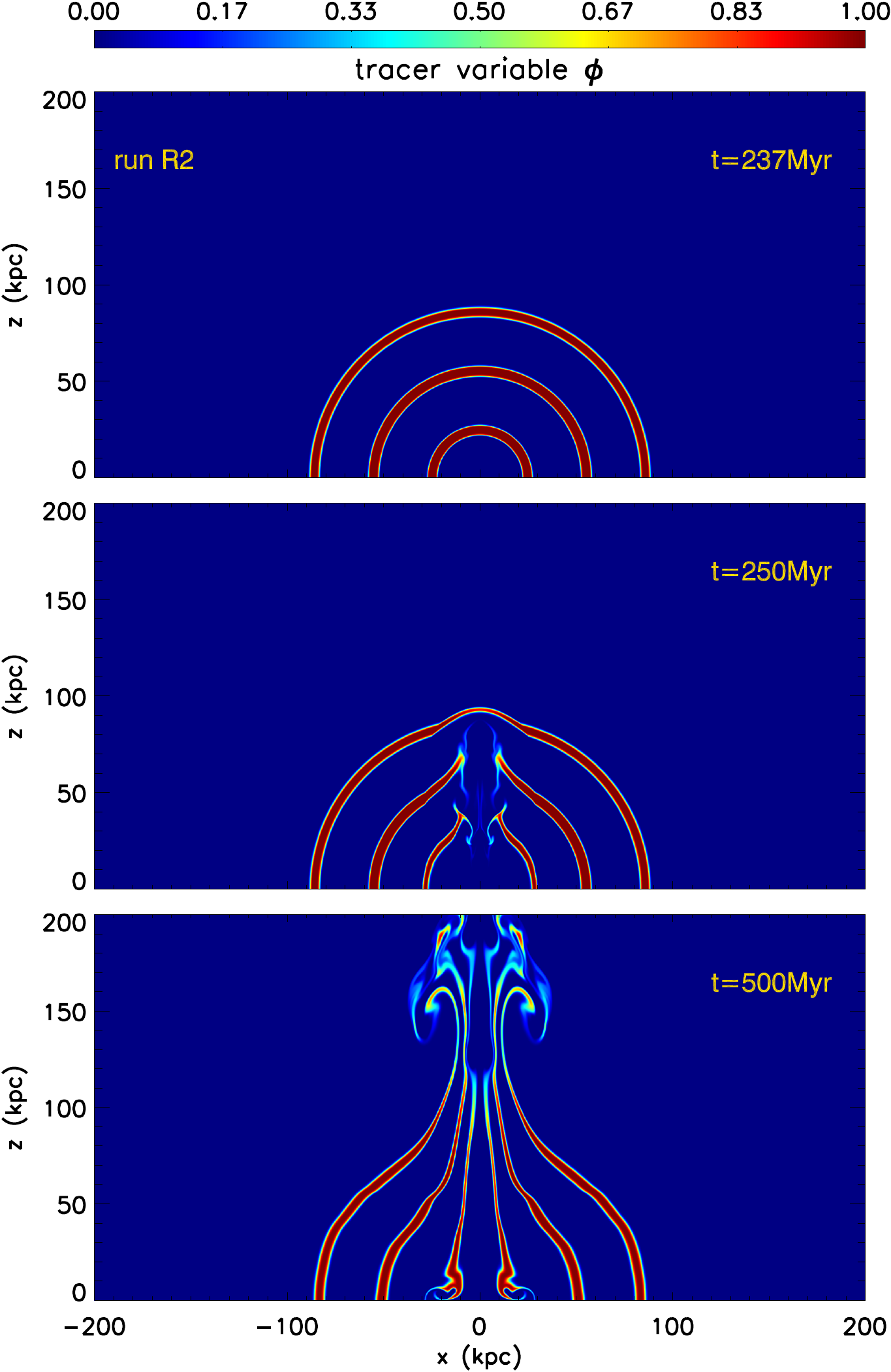}
\caption{Trailing outflows uplifted by AGN bubbles in runs R1 and R2. In both runs, the jets are triggered at $t\sim 238$ Myr when the central cooling catastrophe happens. From left to right: $\it Column$ 1 --- the ICM density in run R1 at three representative times $t=237$, $250$, $500$ Myr, $\it Column$ 2 --- evolution of the density distribution in run R2, and $\it Column$ 3 --- evolution of the tracer variable $\phi$ in run R2. $\phi$ is initially set to be non-vanishing with the value of $1$ only in three concentric shells $r=[25, 30]$, $[55, 60]$, and $[85, 90]$ kpc.} 
\label{plot1}
 \end{figure*}

%Section2.2
\subsection{Simulations}
\label{section2.2}

We performed a suite of three simulations for our default cluster A1795. Starting from hydrostatic equilibrium, the ICM first evolves due to radiative cooling and gravity, and a central cooling catastrophe happens at $t=t_{\rm cc}$ (\citealt{guo14}; G18). We manually turn on an AGN jet event at $t=t_{\rm cc}$, assuming that it is triggered by accretion of cold gas onto the central SMBH (also see \citealt{binney95}; \citealt{gaspari11}; \citealt{li15}; \citealt{yang16}; \citealt{liyuan17}). We implement the jet by applying inflow boundary conditions to a cylindrical nozzle placed at the cluster center, which inject the mass, momentum, and thermal energy fluxes into active zones along the $z$ axis with an opening angle of $0$ degree. For simplicity, the jet is assumed to be steady and uniform at the jet base, lasting for a duration of $5$ Myr. As explained in detail in G18, other jet parameters include the jet density $\rho_{\rm jet}=1.61\times 10^{-26}$ g cm$^{-3}$, energy density $e_{\rm jet}=1.91\times 10^{-9}$ erg cm$^{-3}$, velocity $v_{\rm jet}=3.0\times 10^{9}$ cm s$^{-1}$, and radius $R_{\rm jet}=1.5$ kpc, corresponding to a kinetic-energy-dominated jet with the total kinetic energy of $2.30 \times 10^{60}$ erg, total thermal energy of $6.06 \times 10^{58}$ erg, and the jet power of $1.50 \times 10^{46}$ erg/s. As shown in G18, these jet parameters ensure that the AGN event is powerful enough to avert the cooling flow, while driving a forward shock with aspect ratio consistent with observations.

Table 1 lists some key parameters of our three simulations. Run R1 is a non-viscous simulation with $\mu_{\rm visc}=0$, while in runs R2 and R3, we turn on a relatively low level of constant shear viscosity to suppress the Kelvin-Helmholz instability: $\mu_{\rm visc}=150$ g cm$^{-1}$ s$^{-1}$, which is much lower than the Spitzer value in the hot ICM ($\mu_{\rm visc}\sim 1000$ g cm$^{-1}$ s$^{-1}$ at $T= 5\times10^{7}$ K; \citealt{spitzer62}; \citealt{guo12b}). The initial ICM metallicity distributions in runs R1 and R2 are chosen to be spatially uniform $Z=0.4Z_{\sun}$, while in run R3, we adopt a radius-dependent analytic metallicity profile from \citet{guo10b}, which provides a reasonable fit to the observed iron abundance profile in A1795:
\begin{eqnarray}
Z= (Z^{\beta}_{0}+Z_{r}^{\beta})^{1/\beta} \label{eqnmet}
\end{eqnarray}
%\noindent  
where $\beta = 5$ , $Z_{0} = 0.27Z_{\sun}$ is the average iron abundance observed at large radii, and $Z_{r} = 0.8e^{-r/160 {\rm kpc}}Z_{\sun}$ represents the central metallicity peak often observed in cool core clusters \citep{degrandi04}. Note that metallicity affects the gas cooling rate, and it can also be used to trace ICM motions.

%Section2.3
\subsection{Tracer Variable and Virtual Particles}
\label{section2.3}

In order to track the motions of trailing outflows in a Lagrangian way, we implemented two independent methods in runs R1 and R2. In the first method, we follow the evolution of a scalar tracer variable $\phi$, which is constant along the trajectory of each fluid element: 
\begin{eqnarray}
\frac{d\phi}{dt} = \frac{\partial\phi}{\partial t}+{\bf v} \cdot\nabla \phi = 0.
\end{eqnarray} 
Similar to the metallicity evolution (Eq. \ref{hydro4}), this is implemented in the code in the conservative form (\citealt{saxton01}; \citealt{guo10b}; G18):
\begin{eqnarray}
\frac{\partial (\rho \phi)}{\partial t} + \nabla \cdot ({\rho \phi  \bf v}) = 0.
\end{eqnarray}
For the initial condition at $t=0$, we assume that the value of $\phi$ is $1$ in three concentric shells $r\equiv \sqrt{R^{2}+z^{2}}=[25, 30]$, $[55, 60]$, and $[85, 90]$ kpc, and zero elsewhere. Except for the effect of numerical mixing, the value of $\phi$ for each fluid element remains constant as it moves in our simulations.

In the second method, we directly follow the trajectories of some fluid elements by injecting a collection of ten virtual particles into our simulation domain at $t=0$. These virtual particles are initially located at $R=3$, $6$, $9$, $12$, and $15$ kpc along two lines with $z=15$ and $30$ kpc, respectively, and move according to the local ICM velocity $\partial {\bf r_{\rm i}}/\partial t = {\bf v}$. We have investigated the trajectories of a large collection of virtual particles, and here the trajectories of these ten representative particles are chosen to show the motions of fluid elements in and around the uplifted trailing outflows.

 %%%%%%%%%%%%%%%%%%%%%%%%%%%%%%%%%%%%%%%%%%%%%%%%%%%%%%%%%%%%%%%%%%%%%%%%% 
%Section 3  
\section{Results and Discussions}
\label{section3}

%Section3.1
\subsection{Hot Outflows Uplifted by AGN Bubbles}
\label{section3.1}

%Figure 2
\begin{figure}
\centering
\includegraphics[width=0.45\textwidth]{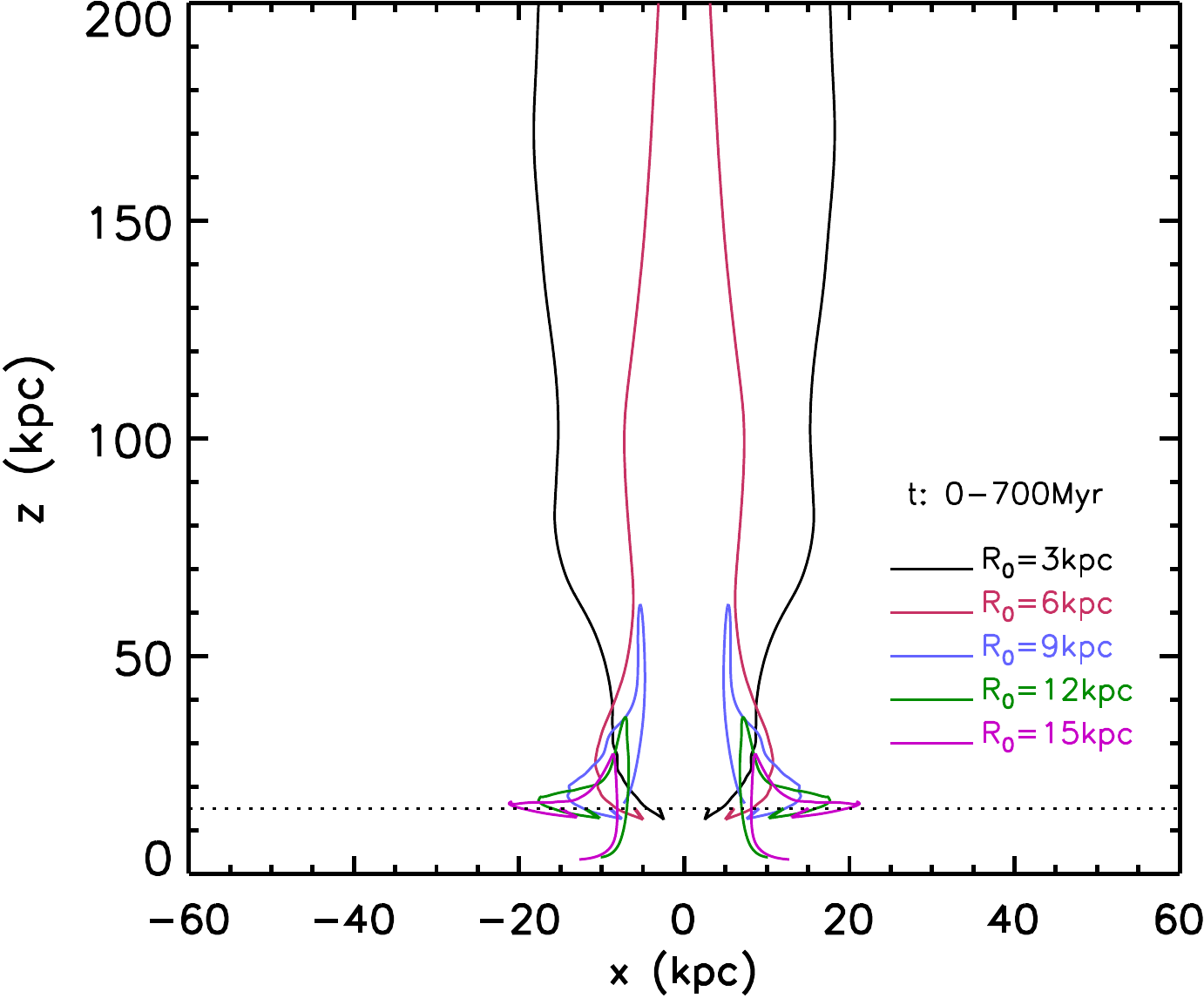} 
\includegraphics[width=0.45\textwidth]{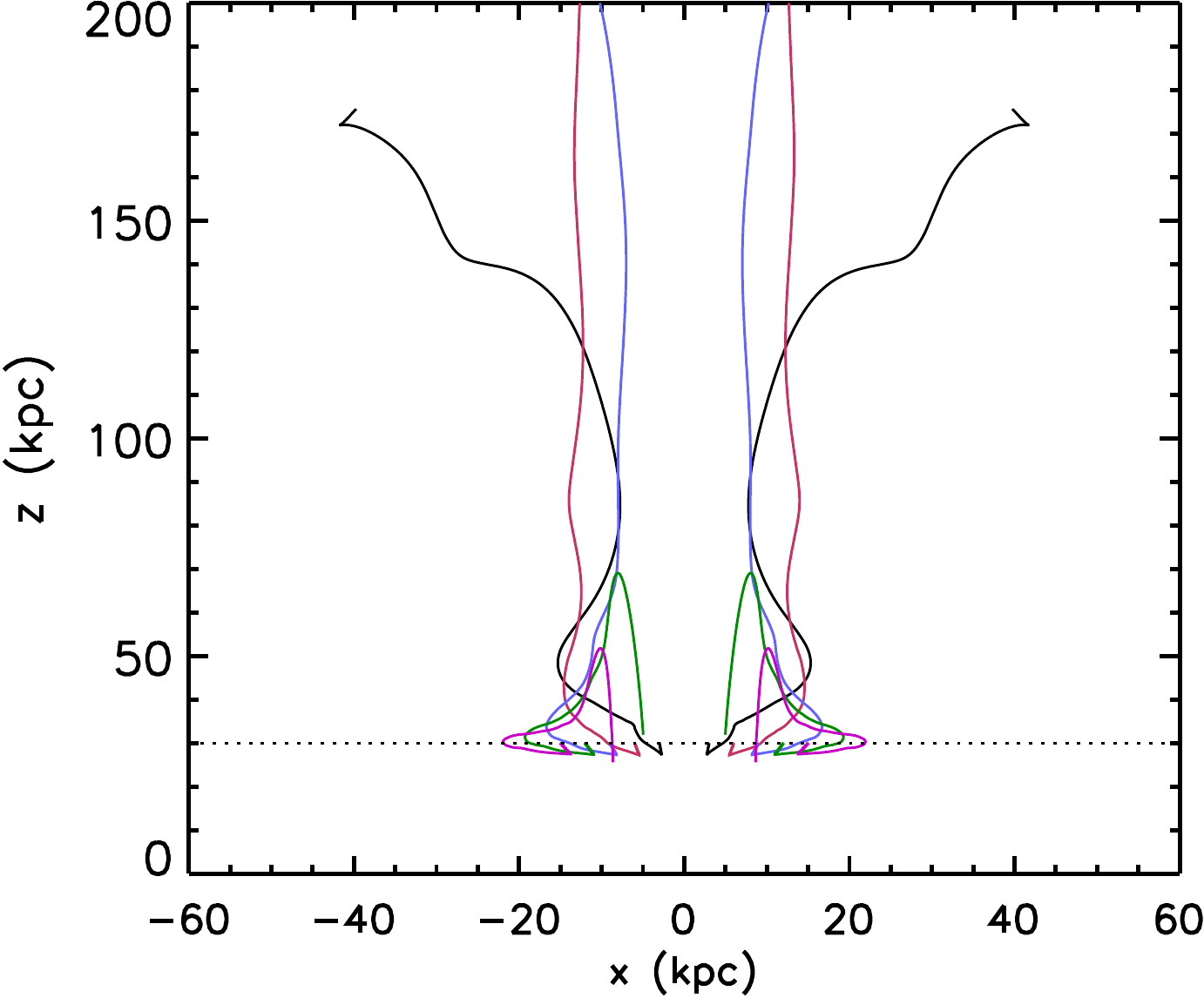}
\caption{Trajectories of some representative virtual particles between $t=0$ and $700$ Myr in Run R2, which help reveal gas motions within trailing outflows. These virtual particles are initially located on the horizontal dotted line in both the top (with $z=15$ kpc) and bottom (with $z=30$ kpc) panels. The x-axes in both panels are stretched horizontally to better show different trajectories. } 
\label{plot2}
 \end{figure}
 
%Figure 3
\begin{figure*}
\centering
\includegraphics[width=0.237\textwidth]{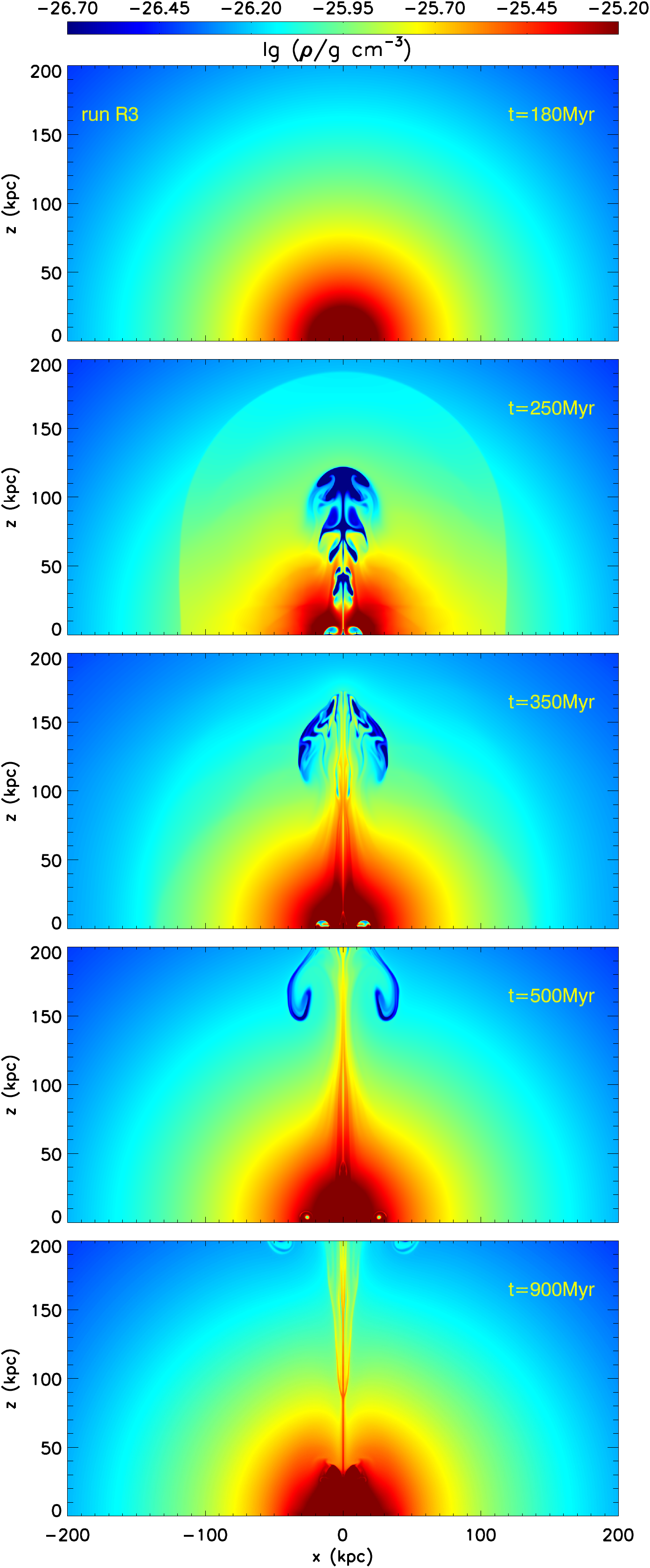} ~~
\includegraphics[width=0.235\textwidth]{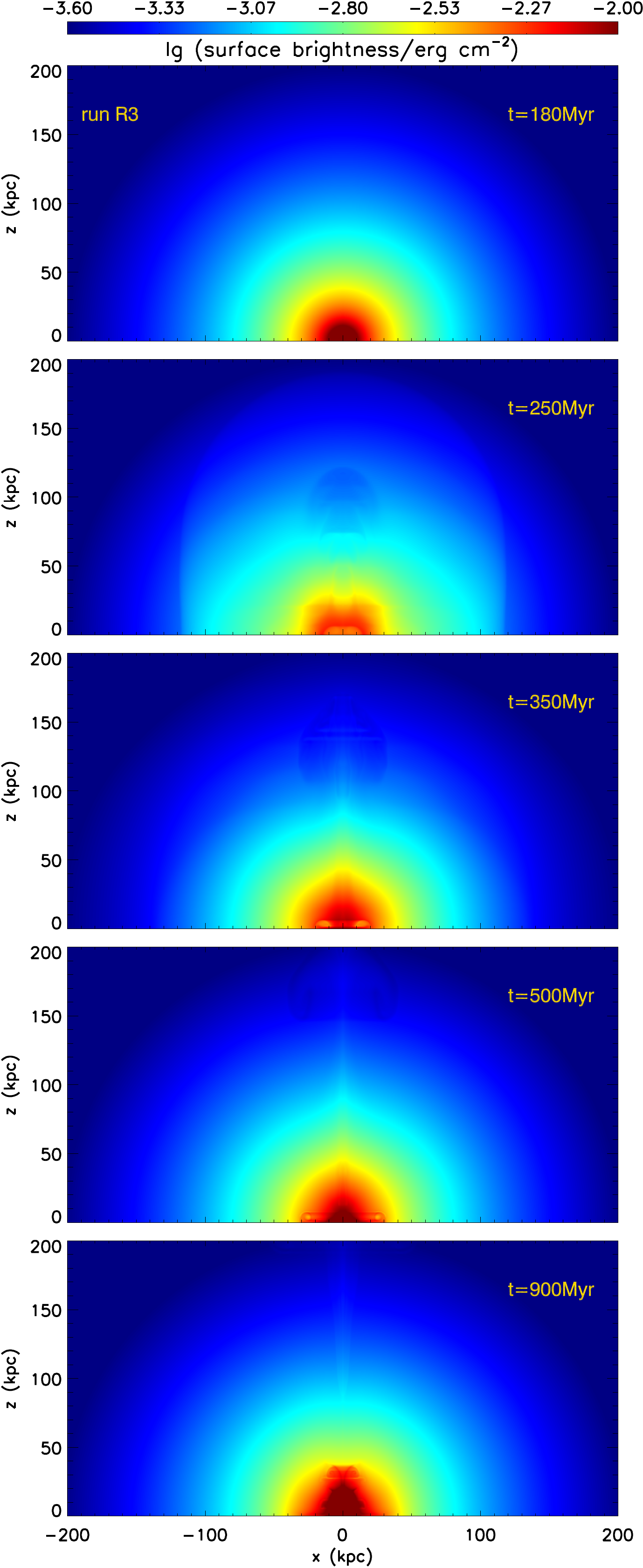}~~
\includegraphics[width=0.232\textwidth]{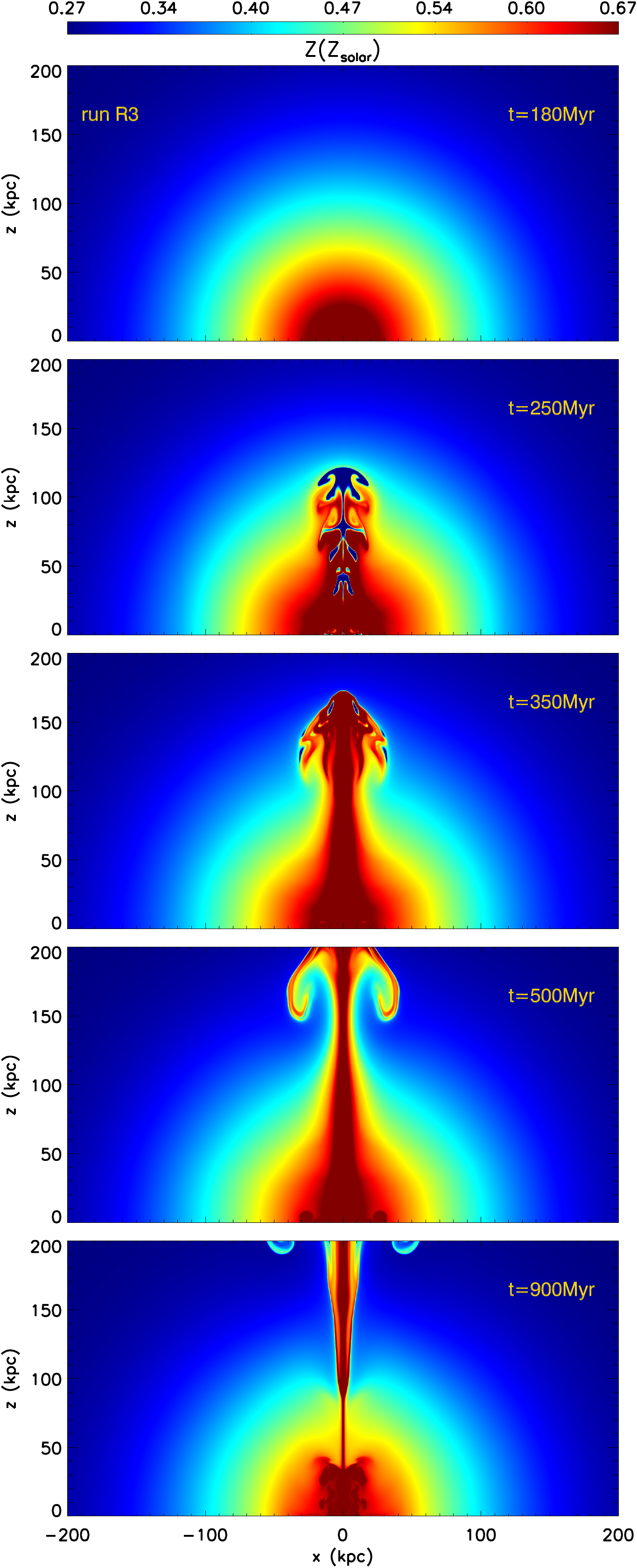}~~
\includegraphics[width=0.232\textwidth]{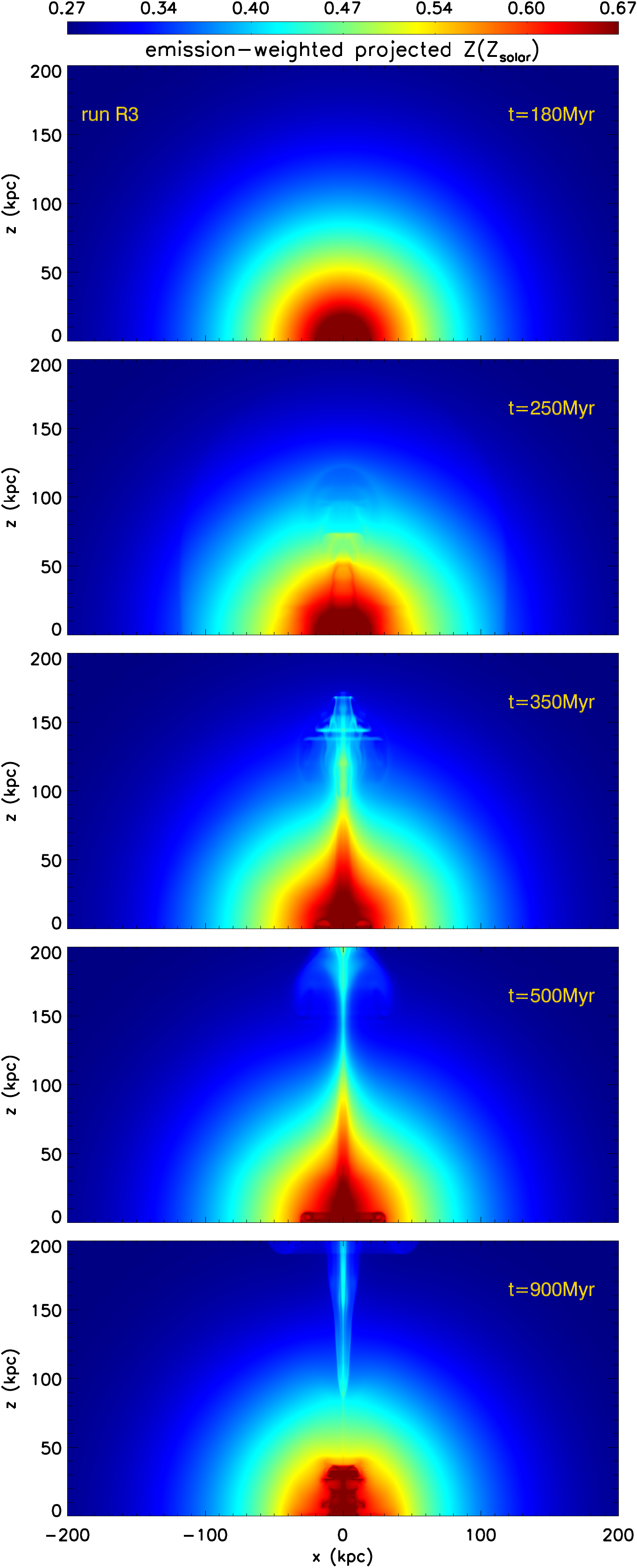}
\caption{Metal-rich trailing outflows uplifted by AGN Bubbles in run R3. From left to right: $\it Column$ 1 --- the evolution of gas density in logarithmic scale, $\it Column$ 2 --- the evolution of synthetic X-ray surface brightness in units of erg cm$^{-2}$ in logarithmic scale (projection of the radiative cooling rate $\mathcal{C}$ perpendicular to the $z$ axis), $\it Column$ 3 --- the evolution of the metallicity $Z (Z_{\sun})$ distribution, and $\it Column$ 4 --- the evolution of the X-ray emission-weighted projected metallicity distribution along a direction perpendicular to the $z$ axis. The time of each row, from top to bottom, is $t=180$, $250$, $350$, $500$, and $900$ Myr, respectively. Note that the jet is turned on at $t\sim180$ Myr when the central cooling catastrophe happens.}
 \label{plot3}
 \end{figure*} 

 %%%%%%%%%%

We first study the formation and evolution of trailing outflows in runs R1 and R2 with the tracer variable and virtual particle methods. Run R1 is a non-viscous simulation, while R2 is a viscous run with $\mu_{\rm visc}=150$ g cm$^{-1}$ s$^{-1}$ where interface instabilities are suppressed significantly. In both runs, the gas metallicity is fixed with $Z = 0.4 Z_{\sun}$, and  AGN jets are manually triggered at $t\sim 238$ Myr (with a duration of $5$ Myr) when the central cooling catastrophe happens (see G18 for more details). 

The left and middle columns of Figure \ref{plot1} show the evolution of the ICM density distribution in logarithmic scale in runs R1 and R2, respectively. Radiative cooling induces gas inflows, leading to gas accumulations in the cluster center, as shown in the top panels, which correspond to $t = 237$ Myr right before the central cooling catastrophe happens. The middle panels indicate that the AGN jet event drives a forward shock, resulting in cool core expansion. At $t = 500$ Myr shown in the bottom panels, the low-density jet plumes (referred as AGN bubbles) are seen at $z>100$ kpc, and a thin over-dense filamentary structure clearly extends from the cluster center to $z\sim 200$ kpc along the $z$ axis. This filament is ``uplifted" by the buoyantly-rising AGN bubble, resembling {\it Darwin Drift} studied in fluid mechanics \citep{darwin53}, and here following G18, we refer to it as ``trailing outflows''. As the resulting AGN bubble in the non-viscous run R1 is strongly disrupted by the KH instability, hereafter we will focus on our viscous runs R2 and R3.

The right $\it column$ of Figure \ref{plot1} shows the evolution of the tracer variable $\phi$ in run R2. At $t=0$, $\phi$ is set to be $1$ in three concentric shells and zero elsewhere (see Section 2.3). Fluid elements in these shells (marked in red in this column) moves following the local gas motion, and the deformation of the marked shells is similar to the evolution of marked planes of {\it Darwin drift} investigated in fluid mechanics (\citealt{dabiri06}; \citealt{pushkin13}; and \citealt{peters16}). The bottom panel in the right column of Figure \ref{plot1} clearly indicates that the ICM gas is indeed physically uplifted to larger distances by the rising AGN bubble, forming real outflows along the original jet direction. Ideally, the tracer variable should maintain constant as fluid elements moves, but numerical diffusion is unavoidable in our grid-based code, particularly in regions close to the jet axis where gas velocities are relatively high (see the right-bottom panel of Fig. \ref{plot1}).

Figure \ref{plot2} shows the trajectories of some representative virtual particles between $t=0$ and $700$ Myr in run R2. These virtual particles are initially located on the horizontal dotted line in both the top (with $z=15$ kpc) and bottom (with $z=30$ kpc) panels. Before the onset of the AGN event, these particles move towards the cluster center in cooling-induced inflows, and the distance moved during this stage is typically several kpc. After the AGN outburst happens, these particles moves outward quickly as swept up by the AGN-induced forward shock. At later times, these particles move upward along the vertical direction, as uplifted by the AGN bubble in trailing outflows. Some particles initially located very close to the jet axis are uplifted to distances even larger than $200$ kpc, while some other particles initially located slightly further away from the jet axis are uplifted roughly by a few tens kpc and then fall back to the cluster center at later times (about $200$ - $300$ Myr after the onset of the AGN event for several representative particles shown in Fig. \ref{plot2}).

%Figure4
\begin{figure*}
\centering
\includegraphics[height=0.85\textwidth]{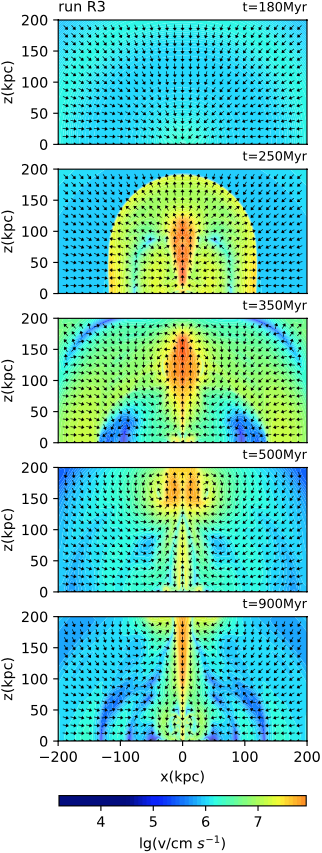} ~~~~
\includegraphics[height=0.85\textwidth]{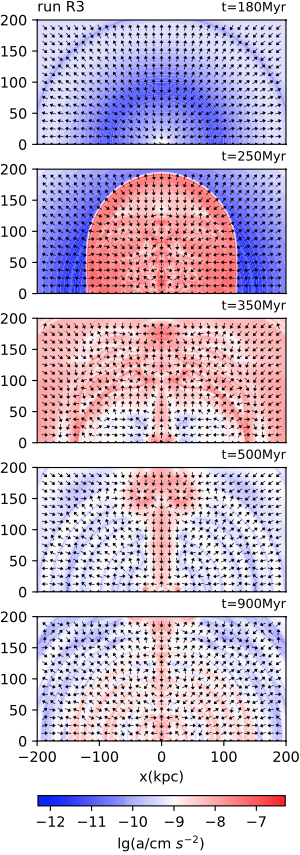}~~~~
\includegraphics[height=0.85\textwidth]{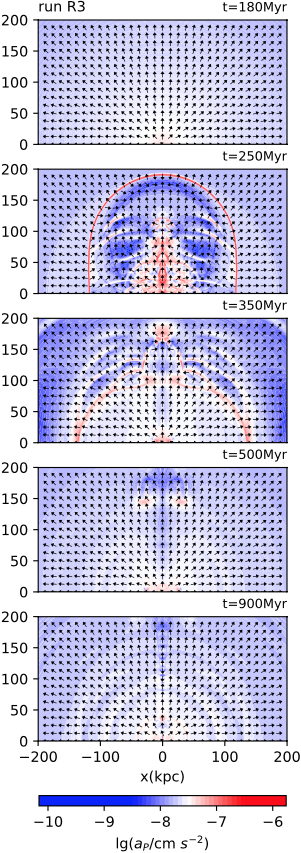}
\caption{Evolution of gas velocity and acceleration in run R3 at the same times as in Figure \ref{plot3}. From left to right: $\it Column$ 1 --- gas velocity, $\it Column$ 2 --- gas acceleration $a=\frac{\partial{\bf v}}{\partial t}+{\bf v} \cdot\nabla {\bf v}$, and $\it Column$ 3 --- acceleration due to gas pressure gradients $ a_{P}=-\frac{1}{\rho}\nabla{P} $. In each column, arrows indicate directions of the corresponding vector, while the background image refers to its magnitude with a color bar shown below}.
 \label{plot4}
 \end{figure*} 

%Figure5
\begin{figure*}
\centering
\includegraphics[width=0.45\textwidth]{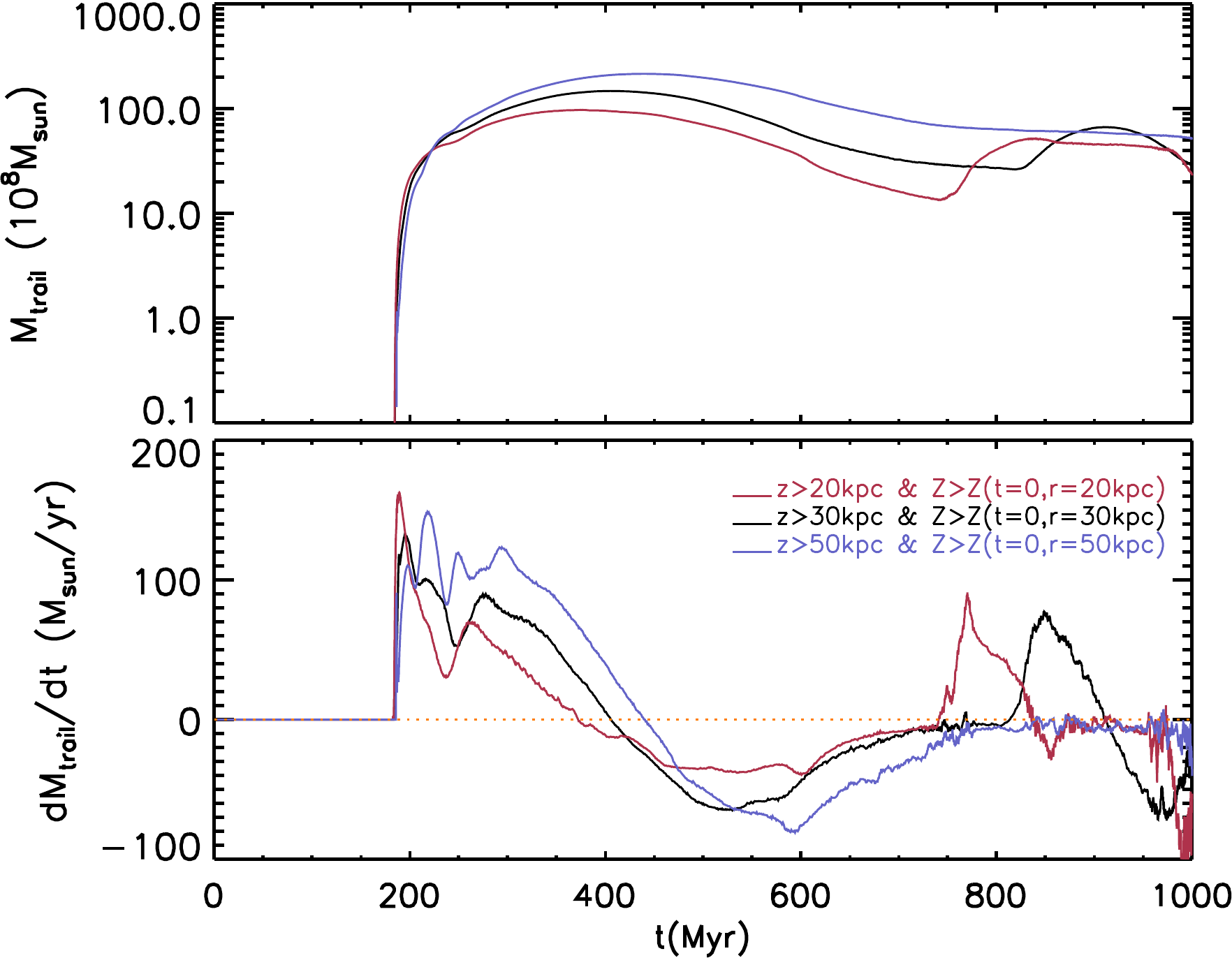} ~~~~~
\includegraphics[width=0.45\textwidth]{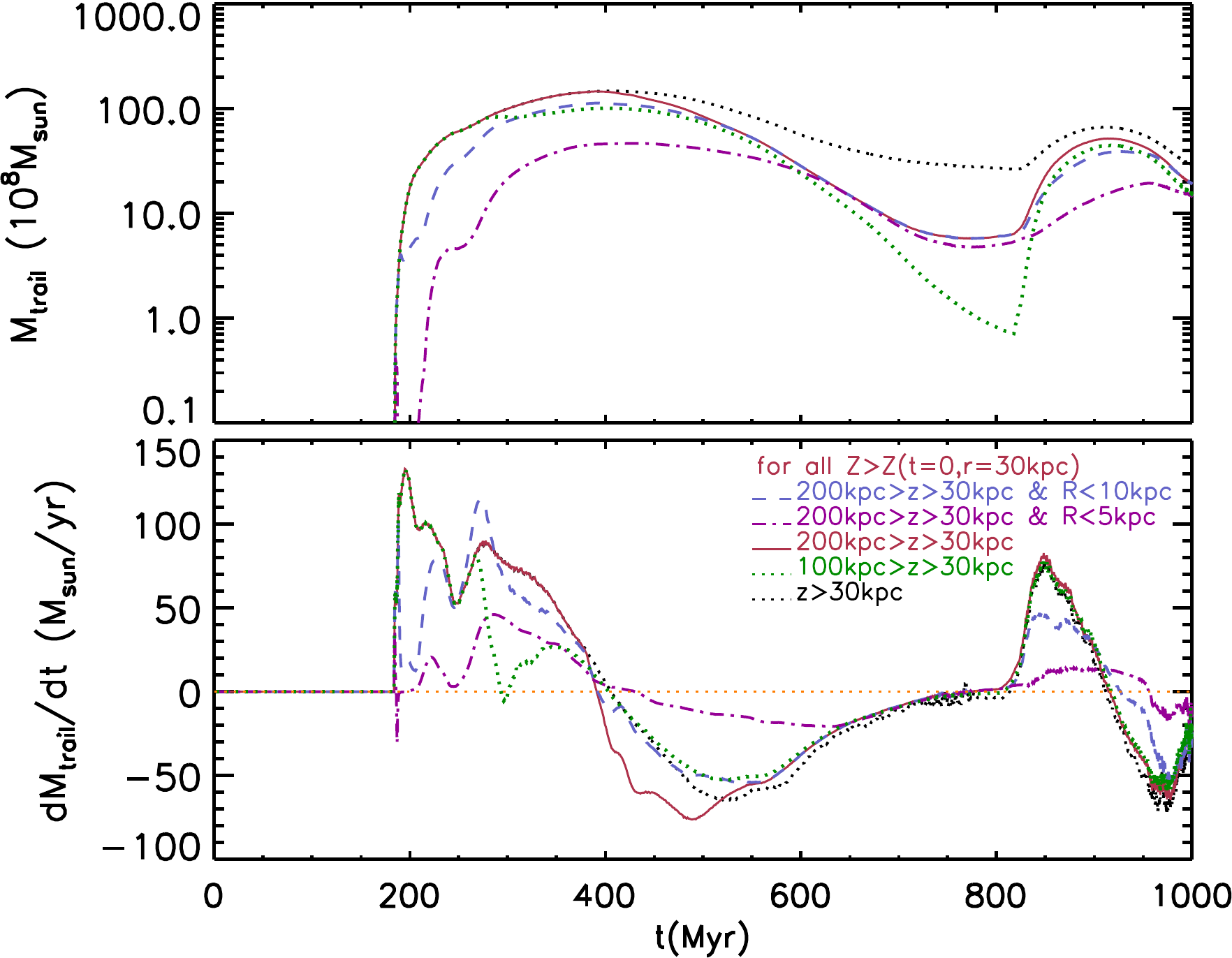}
\caption{Temporal evolution of the gas mass and its change rate of the metal-rich ICM outflow in run R3. {\it Left} --- Evolution of the metal-rich outflow masses (top panel) and rates (bottom panel) across three representative heights $z_{\rm out}=20$ (red), $30$ (black), and $50$ kpc (blue). For each line, the outflow mass across $z=z_{\rm out}$ is estimated by summing up the gas masses in all cells with $z>z_{\rm out}$ and $Z>Z(t=0, r=z_{\rm out})$. {\it Right} --- Evolution of the mass and its change rat of the gas that are uplifted across $z_{\rm out}=30$ kpc and then resides within one of the following spatial regions, including the whole region with $z>30$ kpc (black dotted), the regions with $30 \text{ kpc}<z<100$ kpc (green dotted) and $30 \text{ kpc}<z<200$ kpc (red solid), and two cylindrical regions covering $30<z<200$ kpc with $R<10$ kpc (dashed) and $R<5$ kpc (dot-dashed). The mass change rate has been smoothed over a moving time widow with length of 150 time steps in the simulation. }  
 \label{plot5}
 \end{figure*} 
%%%%%%%%%%%%%
%Section3.2

\subsection{High Metallicity in Trailing Outflows}
\label{section3.2}

{\it Chandra X-ray Observatory} observations of many cool-core clusters show anisotropic metal-rich ICM regions extending preferentially along the axes of X-ray cavities, suggesting that metal-rich hot outflows are driven to large altitudes between twenty to several hundreds kpc by AGN jet events (\citealt{kirp15}; \citealt{kirp11}; \citealt{kirp09}). In our simulations, the  AGN jet event indeed triggers hot ICM outflows from the cluster center to outer regions, as described in Section \ref{section3.1}. Cool-core clusters often have a metallicity peak in central regions enriched presumably by central BCGs, and hot outflows originating from central regions are thus expected to have higher metallicities compared to the ICM gas surrounding them. Adopting a more realistic initial gas metallicity profile (equation \ref{eqnmet}) which contains a central peak, here we investigate metal-rich outflows uplifted by AGN bubbles in more detail in run R3. Due to higher gas metallicities in inner regions compared to runs R1 and R2, radiative cooling in run R3 is more efficient and the central cooling catastrophe happens at an earlier time $t=180$ Myr. We manually turn on AGN jets at this time for a duration of $5$ Myr, and stop the simulation at $t=1$ Gyr.

The ICM evolution in run R3 can be seen clearly in Figure \ref{plot3}, where the columns, from left to right, show the maps of gas density ($\it Column$ 1), synthetic X-ray surface brightness ($\it Column$ 2), metallicity ($\it Column$ 3), and the projected metallicity distribution weighted by X-ray emissivity $\mathcal{C}=n_{\rm i}n_{\rm e}\Lambda$ ($\it Column$ 4). The top row corresponds to $t=180$ Myr, the time right before the onset of the central cooling catastrophe and the AGN jet activity.  At $t=250$ Myr, the second row clearly shows the formation of low-density plumes ($\it Column$ 1) and the X-ray cavity ($\it Column$ 2) resulted from the jet evolution. At $t=350$ Myr in the third row, a drift-like metal-rich filamentary structure (i.e. trailing outflow) is clearly seen extending from the cluster center to the X-ray cavity ($\it Columns$ 1 and 3), while the fourth and fifth rows further show the later evolution of this trailing outflow. It is clear that the observed anisotropic metal-rich ICM structures elongated along X-ray cavities can be naturally produced in our simulation.

To further investigate the formation of trailing outflows, Figure \ref{plot4} shows the maps of gas velocity ($\it Column$ 1), acceleration ($\it Column$ 2), and acceleration contributed by gas pressure gradients ($\it Column$ 3) at the same times as in Figure \ref{plot3}. At $t=180$ Myr, gravity can not be fully balanced by gas pressure gradients due to radiative losses, and the ICM in the whole cluster flows inward towards the cluster center, as seen in the top row of Figure \ref{plot4}. The subsequent AGN jet event induces a forward shock, which propagates through the ICM and reverses the cooling inflow. At $t=250$ Myr, the shock front is located within $r\sim 100$ - $200$ kpc in the second row. The shocked gas first flows outward, and later as the shock front propagates away, flows inward again, as seen in the second and third rows (see G18 for more detail). 

The second and third rows in Figures \ref{plot3} and \ref{plot4} also show that the ICM gas below the jet plumes is being accelerated upward by pressure gradients as the jet plumes move away from the cluster center, forming real metal-rich outflows trailing behind the X-ray cavity similar to {\it Darwin drift} (\citealt{darwin53}; \citealt{pope10}). At $t=500$ Myr, the metal-rich trailing outflow extends from the cluster center to the X-ray cavity ($z\sim 200$ kpc). Note that the ICM gas right above the X-ray cavity is also pushed and accelerated by the rising cavity, forming local meridional circulations around it, as seen in $\it Column$ 1 of Figure \ref{plot4} at $t=350$ and $500$ Myr and discussed in G18. The lower part ($z\lesssim 100$ kpc) of the trailing outflow decelerates at $t=350$ Myr ($\it Column$ 2 of Figure \ref{plot4}), and actually flows back toward the cluster center at $t=500$ Myr ($\it Column$ 1 of Figure \ref{plot4}). At $t=900$ Myr (bottom panels), the whole part of the trailing outflow within $z\lesssim 200$ kpc is flowing inward towards the cluster center ($\it Column$ 1 of Figure \ref{plot4}).     

 Although run R3 is not designed to model a specific AGN event, the properties of our simulated trailing outflows are generally consistent with X-ray observations of hot metal-rich outflows in real clusters (\citealt{kirp15}; \citealt{kirp11}; \citealt{kirp09}). In run R3, metal rich gas in the cluster center is uplifted in trailing outflows of the AGN bubble to a distance of $z\sim 200$ kpc, and the width of the filamentary outflow is $\sim 20$ kpc. The ICM gas entrained in trailing outflows is entirely outflowing at early times, and a growing lower part becomes falling back towards the cluster center at later times (also see G18). In the next subsection, we will further analyze the mass evolution of the ICM gas entrained in trailing outflows.

%Section 3.3
\subsection{Mass Evolution of Trailing Outflows}
\label{section3.3}

Starting with an observationally-motivated radially-declining metallicity profile (Eq. \ref{eqnmet}), run R3 also follows the temporal evolution of the gas metallicity distribution, which serves as a tracer fluid allowing us to estimate the mass evolution of the ICM gas uplifted in trailing outflows. X-ray observations of many galaxy clusters show that the hot gas mass entrained in trailing outflows is typically $10^{9}$ - $10^{10}M_{\odot}$ (about $ 8\times10^{9}M_{\odot}$ in A1795), and the mean outflow rates are typically tens of solar masses per year and upward of $100M_{\odot}$/yr in the extreme (\citealt{kirp15}). Here in this subsection we will see that the outflow mass and mean outflow rate in run R3 are roughly consistent with observations. 

As shown in Figures 1 and 3, trailing outflows are mainly uplifted along the $z$ axis behind the X-ray cavity. To estimate the outflow mass $M_{\rm trail}$ across a specific height $z_{\rm out}$, we sum up the gas masses in all computational cells with $z>z_{\rm out}$ and in which the metallicity $Z$ is higher than the initial metallicity at $r=z_{\rm out}$, i.e., $Z>Z(t=0, r=z_{\rm out})$. In other words, $M_{\rm trail} (z_{\rm out, t})$ represents the total mass of the metal-rich gas that is initially located at $r<z_{\rm out}$, subsequently flows across $z=z_{\rm out}$, and remains residing at $z>z_{\rm out}$ at time $t$. Its temporal change rate $dM_{\rm trail}/dt$ refers to the outflow rate across $z=z_{\rm out}$. The {\it Left} column of Figure \ref{plot5} shows the outflow masses (top panel) and rates (bottom panel) across three representative heights $z_{\rm out}=20$ (red), $30$ (black), and $50$ kpc (blue). The evolutions of outflow masses $M_{\rm trail}$ across these three heights have similar trends during $t_{\rm cc}<t\lesssim 700$ Myr, where $t_{\rm cc}=180$ Myr is the onset time of the jet event (see Table 1). $M_{\rm trail}$ first increases with time as the high-metallicity gas located at $z<z_{\rm out}$ is uplifted in the trailing outflow along the $z$ direction, and at $t\gtrsim 400$ - $500$ Myr, $M_{\rm trail}$ gradually decreases with time due to the falling down of the lower part of the trailing outflow (see the left column in Fig. 4). 

The left panels of Figure 5 also show that at $t\gtrsim 700$ Myr, the values of $M_{\rm trail}$ across $z_{\rm out}=20$ and $30$ kpc increase with time for a duration of about $100$ Myr, which is caused by inner gas circulations triggered after the inflowing lower part of the trailing outflow reaches the cluster center. These inner circulations are limited within $z\lesssim 50$ kpc, as this feature does not show up in the blue line of $M_{\rm trail}$ across $z_{\rm out}=50$ kpc. It is noteworthy that despite of a growing lower part that falls back toward the cluster center at late times, the total gas mass entrained in the trailing outflow still remains to be about several $10^{9}M_{\odot}$ even at the end time of our simulation $t=1$ Gyr.

To further analyze where the uplifted gas is located spatially, we investigate in the right panels of Figure 5 the mass (top) and mass change rate (bottom) of the ICM gas that is uplifted across $z_{\rm out}=30$ kpc and then resides within several spatial regions. In the top right panel of Figure 5, the black dotted line indicates the mass evolution of the total gas uplifted across $z_{\rm out}=30$ kpc, while the green dotted and red solid lines represent the fractions of the uplifted gas residing in $30 \text{ kpc}<z<100$ kpc and $30 \text{ kpc}<z<200$ kpc, respectively. During the first $\sim 100$ Myr after the AGN jet event (i.e., $180<t\lesssim 280$ Myr), these three lines in the right panels (both top and bottom) almost coincide with each other, indicating that the outflow gas mainly resides within $30 \text{ kpc}<z<100$ kpc and increases quickly to $\sim 10^{10}M_{\odot}$ with an outflow rate of nearly $100M_{\odot}$/yr during this stage. At later times, the outflow gas enters into $z>100$ kpc and $z>200$ kpc at $t \sim 280$ and $400$ Myr, respectively, which explains why the green dotted and red solid lines lie below the black dotted line at $t \gtrsim 400$ Myr. The uplifted gas flowing across $z_{\rm out}=30$ kpc clearly forms high-metallicity trailing outflows as seen in Column 3 of Figure 3 at $t=250$ Myr and later times.

The dashed and dot-dashed lines in the {\it right} panels of Figure \ref{plot5} show the mass evolution of the metal-rich trailing outflow uplifted across $z_{\rm out}=30$ kpc and residing 
in two cylinder regions covering $30<z<200$ kpc with $R<10$ kpc and $R<5$ kpc, respectively. The dashed line roughly follows the red solid line, indicating that most outflow gas uplifted across $z_{\rm out}=30$ kpc and residing in $30 < z<200$ kpc is located in a filamentary structure with $R<10$ kpc, as also seen visually in Columns 1 and 3 of Figure 3. Compared to other lines, the dot-dashed line shows a time delay in the mass buildup of the uplifted gas within $R<5$ kpc for about few tens Myr after the AGN jet event, as this region is mainly occupied by the jet plumes during this early stage in our simulation.
 
In summary, Figure 5 indicates that the uplifted gas mass in the trailing outflow in run R3 is several $10^{9}M_{\odot}$ to around $10^{10}M_{\odot}$, and the outflow rate is several tens to about 100 solar masses per year during the outflow phase. We also calculated the uplifted iron mass in the trailing outflow, whose temporal evolution resembles the total uplifted gas mass evolution shown in the top panels of Figure 5. The typical gas metallicity in the trailing outflow is about $0.6Z_{\sun}$ - $0.7Z_{\sun}$, as shown in Column 3 of Figure 3, and the uplifted iron mass is roughly $10^{6}$ - $10^{7}M_{\odot}$. The total outflow mass, rate, and the uplifted iron mass are all consistent with recent X-ray observations of trailing outflows \citep{kirp15}.

 %Section 3.4
\subsection{Comparison with the {\it Darwin Drift} Model}
\label{section3.4}
 
Trailing outflows uplifted by buoyant X-ray cavities belong to a general phenomenon in fluid mechanics known as Darwin drift (\citealt{darwin53}; \citealt{pope10}), which has been investigated extensively with both analytical (\citealt{yih85}; \citealt{pushkin13}) and experimental (\citealt{dabiri06}; \citealt{peters16}) methods. As a solid object moves through an ambient fluid, the {\it Darwin drift} refers to a net displacement of some fluid behind the object along its moving direction (\citealt{darwin53}). The drift phenomenon happens no matter whether the moving object is a solid body or bubble (\citealt{dabiri06}). Here we use the Darwin drift model to estimate the mass of trailing outflows and compare it with our simulation.

In the {\it Darwin drift} model, the drift volume $V_{\rm trail}$ can be estimated as (\citealt{darwin53}; \citealt{dabiri06}; \citealt{pope10}):
 \begin{eqnarray}
V_{\rm trail}=kV_{\rm body},\label{eq12} \rm{ ,}
\end{eqnarray} 
where $V_{\rm body}$ is the volume of the moving body (bubble), and $k$ is a numerical constant with the value of $0.5$ for the case of a moving spherical solid object (\citealt{darwin53}) or $0.72$ for the case of a moving vortex bubble (\citealt{dabiri06}). Here we use $k=0.72$ to estimate the mass of the trailing outflow in run R3. From {\it Column} 1 of Figure \ref{plot3}, the volume of the jet plumes (AGN bubble) $V_{\rm body}$ at $t=250$ Myr may be roughly estimated as half of a sphere with radius of $30$ kpc, resulting in $V_{\rm body}\sim 10^{69}$ cm$^{3}$. At earlier times, the jet plumes may be approximated as a cylinder with radius $R\sim 10$ kpc and height $h\sim 90$ kpc (similar to the cylindrical low-density cavity in run R2 shown in {\it Column} 2 of Figure \ref{plot1} at $t=250$ Myr), which gives a similar value for $V_{\rm body}$. Taking the drift density to be $\sim 5\times10^{-26}$ g cm$^{-3}$ (see {\it Column} 1 of Fig. \ref{plot3}), the mass of the trailing flow at $t=250$ Myr in run R3 is $M_{\rm trail}\sim 10^{10}M_{\odot}$, consistent with our results shown in Section \ref{section3.3}. Alternatively, equation (\ref{eq12}) may be rewritten as $M_{\rm trail}=\eta_{\rm jt}^{-1} kM_{\rm cavity}$, where $\eta_{\rm jt}\sim 0.01$ - $0.03$ is the density ratio of the jet plumes to the trailing outflow, and the mass of the jet plumes may be approximated as the total mass injected by the jet $M_{\rm cavity}\sim 2.6\times10^{8}M_{\odot}$. Thus we get a similar value of $M_{\rm trail}\sim 10^{10}M_{\odot}$ for the mass of the trailing outflow. Note that in our simulations (and likely in real galaxy clusters as well), $M_{\rm trail}$ decreases with time at late times as a part of trailing outflows flow back to the cluster center due to gravity, a factor usually not considered in the traditional {\it Darwin drift} model.

%%%%%%%%%%%%%%%%%%%%%%%%%%%%%%%%%%%%%%%%%%%%%%%%%%%%%%%%%%%%%%%%%%%%%%%%% 
%Section4
\section{Summary}
\label{section4}

Motivated by hot metal-rich outflows observed in recent X-ray observations (\citealt{kirp15}), we perform a suite of three hydrodynamic simulations to investigate trailing outflows uplifted by X-ray cavities in a representative cool core cluster (A1795). In our simulations, we follow the jet evolution in the ICM, the formation and evolution of X-ray cavities and trailing outflows in the wakes of X-ray cavities self-consistently. 

To track the motions of trailing outflows, we adopt the tracer variable and virtual particle methods in runs R1 and R2, and show that some of the ICM gas originally located near the cluster center is indeed physically uplifted behind the X-ray cavity, forming a thin filamentary outflow extending from the cluster center to the cavity along the cavity axis up to $z\sim 100$ - $200$ kpc. In run R3, we additionally follow the evolution of an observationally-motivated radially-declining metallicity profile, and show that the hot trailing outflow uplifted from central regions is indeed metal-rich compared to the surrounding ICM, consistent with X-ray observations. The gas entrained in the trailing outflow is entirely outflowing at early times, and as the cavity rises further away from the cluster center at later times, a growing lower part of the trailing outflow becomes falling back towards the cluster center.

In the R3 simulation, the gas mass in the trailing outflow rises quickly to around $10^{10}M_{\odot}$ within about $100$ Myr with an average outflow rate of nearly $100M_{\odot}$/yr. The uplifted iron mass is about $10^{6}$ - $10^{7}M_{\odot}$, consistent with observations (\citealt{kirp15}). Physically, trailing outflows belong to the phenomenon of {\it Darwin Drift} previously studied in fluid mechanics, and the outflow mass in our simulation is consistent with that estimated in the drift model. At later times, the outflow mass decreases gradually as a lower part of the trailing outflow falls back towards the cluster center due to gravity. However, even at the end of our simulation $t=1$ Gyr (much longer than the expected duration between two consecutive AGN jet events in real clusters), the total mass of uplifted high-metallicity gas still remains to be about several $10^{9}M_{\odot}$, which may contribute to the enrichment of the bulk ICM and the broadening of central metallicity peaks observed in cool core clusters. The latter has been previously attributed to diffusive transport of metals by stochastic gas motions in the ICM (e.g., \citealt{rebusco05}).

Trailing outflows (drifts) uplifted by X-ray cavities are one type of outflows driven by AGN outbursts, and naturally explain hot metal-rich outflows observed in galaxy groups and clusters, which tend to align with the large-scale cavity axes. Along the axes of trailing outflows, gas density is relatively high, leading to efficient radiative cooling, and cold gas may drop out due to local thermal instability, forming cool filaments observed in some galaxy clusters (e.g., \citealt{mcnamara14}; \citealt{tremblay15}). However, in our simulations R1-R3, cold gas with temperatures below $5 \times 10^{5}$ K only appears near the cluster center and cool extended filaments do not form, suggesting that a much higher spatial resolution may be required to study local thermal instability in trailing outflows (\citealt{revaz08}). We leave a careful investigation of the formation of cool filaments in trailing outflows to future studies.

%%%%%%%%%%%%%%%%%%%%%%%%%%%%%%%%%%%%%%%%%%%%%%%%%%%%%%%%%%%%%%%%%%%%%%%%% 

\acknowledgements 

We are grateful to an anonymous referee for helpful comments and suggestions. This work was supported in part by Chinese Academy of Sciences through the Hundred Talents Program and the Key Research Program of Frontier Sciences (No. QYZDB-SSW-SYS033 and QYZDJ-SSW-SYS008), Natural Science Foundation of China (No. 11633006), and Natural Science Foundation of Shanghai (No. 18ZR1447100). The simulations presented in this work were performed using the high performance computing resources in the Core Facility for Advanced Research Computing at Shanghai Astronomical Observatory.

\bibliography{ms}

\begin{thebibliography}{47}
\expandafter\ifx\csname natexlab\endcsname\relax\def\natexlab#1{#1}\fi

\bibitem[{{Anderson} {et~al.}(2018){Anderson}, {Gaensler}, {Heald},
  {O'Sullivan}, {Kaczmarek}, \& {Feain}}]{anderson18}
{Anderson}, C.~S., {Gaensler}, B.~M., {Heald}, G.~H., {O'Sullivan}, S.~P.,
  {Kaczmarek}, J.~F., \& {Feain}, I.~J. 2018, \apj, 855, 41

\bibitem[{{Binney} \& {Tabor}(1995)}]{binney95}
{Binney}, J., \& {Tabor}, G. 1995, \mnras, 276, 663

\bibitem[{{Br{\"u}ggen}(2003)}]{bruggen03}
{Br{\"u}ggen}, M. 2003, \apj, 592, 839

\bibitem[{{Churazov} {et~al.}(2001){Churazov}, {Br{\"u}ggen}, {Kaiser},
  {B{\"o}hringer}, \& {Forman}}]{churazov01}
{Churazov}, E., {Br{\"u}ggen}, M., {Kaiser}, C.~R., {B{\"o}hringer}, H., \&
  {Forman}, W. 2001, \apj, 554, 261

\bibitem[{{Dabiri}(2006)}]{dabiri06}
{Dabiri}, J.~O. 2006, Journal of Fluid Mechanics, 547, 105

\bibitem[{{Darwin}(1953)}]{darwin53}
{Darwin}, C. 1953, Proceedings of the Cambridge Philosophical Society, 49, 342

\bibitem[{{De Grandi} {et~al.}(2004){De Grandi}, {Ettori}, {Longhetti}, \&
  {Molendi}}]{degrandi04}
{De Grandi}, S., {Ettori}, S., {Longhetti}, M., \& {Molendi}, S. 2004, \aap,
  419, 7

\bibitem[{{Fabian} {et~al.}(2003){Fabian}, {Sanders}, {Crawford}, {Conselice},
  {Gallagher}, \& {Wyse}}]{fabian03}
{Fabian}, A.~C., {Sanders}, J.~S., {Crawford}, C.~S., {Conselice}, C.~J.,
  {Gallagher}, J.~S., \& {Wyse}, R.~F.~G. 2003, \mnras, 344, L48

\bibitem[{{Gaspari} {et~al.}(2011){Gaspari}, {Melioli}, {Brighenti}, \&
  {D'Ercole}}]{gaspari11}
{Gaspari}, M., {Melioli}, C., {Brighenti}, F., \& {D'Ercole}, A. 2011, \mnras,
  411, 349

\bibitem[{{Guo}(2015)}]{guo15}
{Guo}, F. 2015, \apj, 803, 48

\bibitem[{{Guo}(2016)}]{guo16}
---. 2016, \apj, 826, 17

\bibitem[{{Guo} {et~al.}(2018){Guo}, {Duan}, \& {Yuan}}]{guo18}
{Guo}, F., {Duan}, X., \& {Yuan}, Y.-F. 2018, \mnras, 473, 1332

\bibitem[{{Guo} \& {Mathews}(2010{\natexlab{a}})}]{guo10b}
{Guo}, F., \& {Mathews}, W.~G. 2010{\natexlab{a}}, \apj, 717, 937

\bibitem[{{Guo} \& {Mathews}(2010{\natexlab{b}})}]{guo10a}
---. 2010{\natexlab{b}}, \apj, 712, 1311

\bibitem[{{Guo} \& {Mathews}(2011)}]{guo11}
---. 2011, \apj, 728, 121

\bibitem[{{Guo} \& {Mathews}(2012)}]{guo12}
---. 2012, \apj, 756, 181

\bibitem[{{Guo} \& {Mathews}(2014)}]{guo14}
---. 2014, \apj, 780, 126

\bibitem[{{Guo} {et~al.}(2012){Guo}, {Mathews}, {Dobler}, \& {Oh}}]{guo12b}
{Guo}, F., {Mathews}, W.~G., {Dobler}, G., \& {Oh}, S.~P. 2012, \apj, 756, 182

\bibitem[{{Hatch} {et~al.}(2006){Hatch}, {Crawford}, {Johnstone}, \&
  {Fabian}}]{hatch06}
{Hatch}, N.~A., {Crawford}, C.~S., {Johnstone}, R.~M., \& {Fabian}, A.~C. 2006,
  \mnras, 367, 433

\bibitem[{{Kirkpatrick} {et~al.}(2009){Kirkpatrick}, {Gitti}, {Cavagnolo},
  {McNamara}, {David}, {Nulsen}, \& {Wise}}]{kirp09}
{Kirkpatrick}, C.~C., {Gitti}, M., {Cavagnolo}, K.~W., {McNamara}, B.~R.,
  {David}, L.~P., {Nulsen}, P.~E.~J., \& {Wise}, M.~W. 2009, \apjl, 707, L69

\bibitem[{{Kirkpatrick} \& {McNamara}(2015)}]{kirp15}
{Kirkpatrick}, C.~C., \& {McNamara}, B.~R. 2015, \mnras, 452, 4361

\bibitem[{{Kirkpatrick} {et~al.}(2011){Kirkpatrick}, {McNamara}, \&
  {Cavagnolo}}]{kirp11}
{Kirkpatrick}, C.~C., {McNamara}, B.~R., \& {Cavagnolo}, K.~W. 2011, \apjl,
  731, L23

\bibitem[{{Li} {et~al.}(2015){Li}, {Bryan}, {Ruszkowski}, {Voit}, {O'Shea}, \&
  {Donahue}}]{li15}
{Li}, Y., {Bryan}, G.~L., {Ruszkowski}, M., {Voit}, G.~M., {O'Shea}, B.~W., \&
  {Donahue}, M. 2015, \apj, 811, 73

\bibitem[{{Li} {et~al.}(2016){Li}, {Ruszkowski}, \& {Bryan}}]{liyuan17}
{Li}, Y., {Ruszkowski}, M., \& {Bryan}, G.~L. 2016, ArXiv: 1611.05455

\bibitem[{{McNamara} {et~al.}(2014){McNamara}, {Russell}, {Nulsen}, {Edge},
  {Murray}, {Main}, {Vantyghem}, {Combes}, {Fabian}, {Salome}, {Kirkpatrick},
  {Baum}, {Bregman}, {Donahue}, {Egami}, {Hamer}, {O'Dea}, {Oonk}, {Tremblay},
  \& {Voit}}]{mcnamara14}
{McNamara}, B.~R., {et~al.} 2014, \apj, 785, 44

\bibitem[{{McNamara} {et~al.}(2016){McNamara}, {Russell}, {Nulsen}, {Hogan},
  {Fabian}, {Pulido}, \& {Edge}}]{mcnamara16}
{McNamara}, B.~R., {Russell}, H.~R., {Nulsen}, P.~E.~J., {Hogan}, M.~T.,
  {Fabian}, A.~C., {Pulido}, F., \& {Edge}, A.~C. 2016, \apj, 830, 79

\bibitem[{{O'Sullivan} {et~al.}(2013){O'Sullivan}, {Feain},
  {McClure-Griffiths}, {Ekers}, {Carretti}, {Robishaw}, {Mao}, {Gaensler},
  {Bland-Hawthorn}, \& {Stawarz}}]{sullivan13}
{O'Sullivan}, S.~P., {et~al.} 2013, \apj, 764, 162

\bibitem[{{Peters} {et~al.}(2016){Peters}, {Madonia}, {Lohse}, \& {van der
  Meer}}]{peters16}
{Peters}, I.~R., {Madonia}, M., {Lohse}, D., \& {van der Meer}, D. 2016, ArXiv
  e-prints, arXiv:1601.03078

\bibitem[{{Pope} {et~al.}(2010){Pope}, {Babul}, {Pavlovski}, {Bower}, \&
  {Dotter}}]{pope10}
{Pope}, E.~C.~D., {Babul}, A., {Pavlovski}, G., {Bower}, R.~G., \& {Dotter}, A.
  2010, \mnras, 406, 2023

\bibitem[{{Pushkin} {et~al.}(2013){Pushkin}, {Shum}, \& {Yeomans}}]{pushkin13}
{Pushkin}, D.~O., {Shum}, H., \& {Yeomans}, J.~M. 2013, Journal of Fluid
  Mechanics, 726, 5

\bibitem[{{Rebusco} {et~al.}(2005){Rebusco}, {Churazov}, {B{\"o}hringer}, \&
  {Forman}}]{rebusco05}
{Rebusco}, P., {Churazov}, E., {B{\"o}hringer}, H., \& {Forman}, W. 2005,
  \mnras, 359, 1041

\bibitem[{{Revaz} {et~al.}(2008){Revaz}, {Combes}, \& {Salom{\'e}}}]{revaz08}
{Revaz}, Y., {Combes}, F., \& {Salom{\'e}}, P. 2008, \aap, 477, L33

\bibitem[{{Reynolds} {et~al.}(2005){Reynolds}, {McKernan}, {Fabian}, {Stone},
  \& {Vernaleo}}]{reynolds05}
{Reynolds}, C.~S., {McKernan}, B., {Fabian}, A.~C., {Stone}, J.~M., \&
  {Vernaleo}, J.~C. 2005, \mnras, 357, 242

\bibitem[{{Roediger} {et~al.}(2007){Roediger}, {Br{\"u}ggen}, {Rebusco},
  {B{\"o}hringer}, \& {Churazov}}]{roediger07}
{Roediger}, E., {Br{\"u}ggen}, M., {Rebusco}, P., {B{\"o}hringer}, H., \&
  {Churazov}, E. 2007, \mnras, 375, 15

\bibitem[{{Russell} {et~al.}(2017){Russell}, {McNamara}, {Fabian}, {Nulsen},
  {Combes}, {Edge}, {Hogan}, {McDonald}, {Salom{\'e}}, {Tremblay}, \&
  {Vantyghem}}]{russell17}
{Russell}, H.~R., {et~al.} 2017, \mnras, 472, 4024

\bibitem[{{Salom{\'e}} {et~al.}(2006){Salom{\'e}}, {Combes}, {Edge},
  {Crawford}, {Erlund}, {Fabian}, {Hatch}, {Johnstone}, {Sanders}, \&
  {Wilman}}]{salom06}
{Salom{\'e}}, P., {et~al.} 2006, \aap, 454, 437

\bibitem[{{Salom{\'e}} {et~al.}(2008){Salom{\'e}}, {Revaz}, {Combes}, {Pety},
  {Downes}, {Edge}, \& {Fabian}}]{salom08}
{Salom{\'e}}, P., {Revaz}, Y., {Combes}, F., {Pety}, J., {Downes}, D., {Edge},
  A.~C., \& {Fabian}, A.~C. 2008, \aap, 483, 793

\bibitem[{{Saxton} {et~al.}(2001){Saxton}, {Sutherland}, \&
  {Bicknell}}]{saxton01}
{Saxton}, C.~J., {Sutherland}, R.~S., \& {Bicknell}, G.~V. 2001, \apj, 563, 103

\bibitem[{{Simionescu} {et~al.}(2009){Simionescu}, {Werner}, {B{\"o}hringer},
  {Kaastra}, {Finoguenov}, {Br{\"u}ggen}, \& {Nulsen}}]{simionescu09}
{Simionescu}, A., {Werner}, N., {B{\"o}hringer}, H., {Kaastra}, J.~S.,
  {Finoguenov}, A., {Br{\"u}ggen}, M., \& {Nulsen}, P.~E.~J. 2009, \aap, 493,
  409

\bibitem[{{Spitzer}(1962)}]{spitzer62}
{Spitzer}, L. 1962, {Physics of Fully Ionized Gases, 2nd edition,}
  (Interscience, New York)

\bibitem[{{Stone} \& {Norman}(1992)}]{stone92}
{Stone}, J.~M., \& {Norman}, M.~L. 1992, \apjs, 80, 753

\bibitem[{{Sutherland} \& {Dopita}(1993)}]{sd93}
{Sutherland}, R.~S., \& {Dopita}, M.~A. 1993, \apjs, 88, 253

\bibitem[{{Tremblay} {et~al.}(2015){Tremblay}, {O'Dea}, {Baum}, {Mittal},
  {McDonald}, {Combes}, {Li}, {McNamara}, {Bremer}, {Clarke}, {Donahue},
  {Edge}, {Fabian}, {Hamer}, {Hogan}, {Oonk}, {Quillen}, {Sanders},
  {Salom{\'e}}, \& {Voit}}]{tremblay15}
{Tremblay}, G.~R., {et~al.} 2015, \mnras, 451, 3768

\bibitem[{{Vantyghem} {et~al.}(2016){Vantyghem}, {McNamara}, {Russell},
  {Hogan}, {Edge}, {Nulsen}, {Fabian}, {Combes}, {Salom{\'e}}, {Baum},
  {Donahue}, {Main}, {Murray}, {O'Connell}, {O'Dea}, {Oonk}, {Parrish},
  {Sanders}, {Tremblay}, \& {Voit}}]{vantyghem16}
{Vantyghem}, A.~N., {et~al.} 2016, \apj, 832, 148

\bibitem[{{Vikhlinin} {et~al.}(2006){Vikhlinin}, {Kravtsov}, {Forman}, {Jones},
  {Markevitch}, {Murray}, \& {Van Speybroeck}}]{vikhlinin06}
{Vikhlinin}, A., {Kravtsov}, A., {Forman}, W., {Jones}, C., {Markevitch}, M.,
  {Murray}, S.~S., \& {Van Speybroeck}, L. 2006, \apj, 640, 691

\bibitem[{{Yang} \& {Reynolds}(2016)}]{yang16}
{Yang}, H.-Y.~K., \& {Reynolds}, C.~S. 2016, \apj, 818, 181

\bibitem[{{Yih}(1985)}]{yih85}
{Yih}, C.-S. 1985, Journal of Fluid Mechanics, 152, 163

\end{thebibliography}

\end{document}